%% file: main.tex
\documentclass{article}
\usepackage[cmex10]{amsmath}
\usepackage{amssymb}
\usepackage{pifont} 
\usepackage{tikz}
\usepackage{cite}
\usepackage{flushend}
\usepackage{multirow}
\usepackage{xcolor,colortbl}
\usepackage{arydshln}
\usepackage{graphicx}

\usepackage{algorithm}
\usepackage{algorithmic}

\usepackage[hyphens]{url}

\usepackage{algorithmic}

\usepackage{adjustbox}
\usepackage{booktabs}
\usepackage[affil-it]{authblk}
\newcommand*\circled[1]{\tikz[baseline=(char.base)]{
            \node[shape=circle,draw,inner sep=0.8pt] (char) {#1};}}
\usepackage[capitalise]{cleveref}
\begin{document}
%
\include{symbols.tex}

\title{Unified Power Flow Model for Bipolar HVDC Grids in Unbalanced Operation}

\author[1,2]{Chandra~Kant~Jat}
\author[1,2]{Francesco~Giacomo~Puricelli}
\author[1,2]{Jef~Beerten}
\author[1,2]{Hakan Ergun}
\author[1,2]{Dirk Van Hertem}

\affil[1]{{Department of Electrical Engineering, KU Leuven, Leuven, Belgium}}
\affil[2]{{Etch-EnergyVille, Genk, Belgium}}

\date{}

\maketitle

\begin{abstract}
In the near future, point-to-point High Voltage Direct Current (HVDC) systems are expected to evolve into multi-terminal and meshed HVDC grids, predominantly adopting a bipolar HVDC configuration. Normally, bipolar HVDC systems operate in balanced mode, i.e., near zero current flows through metallic or ground return. However, bipolar HVDC systems can also be operated in an unbalanced mode in case of a single converter pole or line conductor outage. A steady-state analysis of the unbalanced DC network requires solving a power flow problem including various converter control modes, as the steady-state behavior of the converters is governed by their control modes. This paper presents a comprehensive and unified power flow model for the balanced and unbalanced operation of bipolar HVDC grids, including various converter control modes on the AC and DC sides of the converter, in a hybrid AC/DC system. It extends the basic control modes, developed for monopolar HVDC grids, to support the balanced as well as unbalanced operation of bipolar HVDC grids.
Additionally, an AC-droop control, which defines a droop relationship between voltage magnitude and reactive power at the AC side of the converter, is incorporated into the modeling of bipolar HVDC systems. The functionality of the proposed model is demonstrated through a test case, and the power flow results are validated using PSCAD simulations. The impact of converter control modes on post-contingency system states is also investigated for single-pole contingencies.  The proposed model is implemented as an open-source tool in the Julia/JuMP framework, where larger test cases demonstrate the robustness of the model and tool. 

\end{abstract}


\input{full_paper.tex}

\bibliographystyle{IEEEtran}
\bibliography{References_pf}

\end{document}

%% file: symbols.tex
%superscripts
\newcommand{\convs}{cv}
\newcommand{\acs}{ac}
\newcommand{\dcs}{dc}
\newcommand{\transfs}{tf}
\newcommand{\reactor}{pr}
\newcommand{\filter}{f}

%% loads and generators
\newcommand{\Pg}{P_{g}}
\newcommand{\Qg}{Q_{g}}
\newcommand{\Pgmax}{P_{g}^{max}}
\newcommand{\Pgmin}{P_{g}^{min}}
\newcommand{\Qgmax}{Q_{g}^{max}}
\newcommand{\Qgmin}{Q_{g}^{min}}

% topological sets
\newcommand{\TopoDC}{\mathcal{T}^{\text{\dcs}}}
\newcommand{\TopoDCrev}{\mathcal{T}^{\text{\dcs,rev}}}
\newcommand{\TopoAC}{\mathcal{T}^{\text{\acs}}}
\newcommand{\Topoconv}{\mathcal{T}^{\text{\convs}}}
\newcommand{\Topoconvrev}{\mathcal{T}^{\text{\convs,rev}}}
\newcommand{\TopogenAC}{\mathcal{T}^{\text{gen,\acs}}}
\newcommand{\TopoloadAC}{\mathcal{T}^{\text{load,\acs}}}
\newcommand{\TopoloadDC}{\mathcal{T}^{\text{load,\dcs}}}

% AC lines
\newcommand{\Plij}{P^{\text{ac}}_{lij}}
\newcommand{\Plji}{P^{\text{ac}}_{lji}}
\newcommand{\Qlij}{Q^{\text{ac}}_{lij}}
\newcommand{\Qlji}{Q^{\text{ac}}_{lji}}

% AC nodes
\newcommand{\Ui}{U_{i}}
\newcommand{\Umagi}{U^{\text{mag}}_{i}}
\newcommand{\thetai}{\theta_{i}}
\newcommand{\Uj}{U_{j}}
\newcommand{\Umagj}{U^{\text{mag}}_{j}}
\newcommand{\thetaj}{\theta_{j}}
\newcommand{\Uimax}{U^{\text{max}}_{i}}
\newcommand{\Uimin}{U^{\text{min}}_{i}}

% Convex relax
\newcommand{\Wii}{W_{i}}
\newcommand{\Wjj}{W_{j}}
\newcommand{\Rtfij}{R^{\text{\transfs}}_{ic}}
\newcommand{\Ttfij}{T^{\text{\transfs}}_{ic}}
\newcommand{\itfc}{i^{\text{sq,\transfs}}_{c}}
\newcommand{\iprc}{i^{\text{sq,\reactor}}_{c}}

\newcommand{\Wfiltmagi}{W^{\text{\filter}}_{c}}
\newcommand{\Wconvmagi}{W^{\text{\convs}}_{c}}
\newcommand{\Uconvmagimax}{U^{\text{\convs,max}}_{c}}
\newcommand{\Uconvmagimin}{U^{\text{\convs,min}}_{c}}

\newcommand{\Rprij}{R^{\text{\reactor}}_{c}}
\newcommand{\Tprij}{T^{\text{\reactor}}_{c}}

%%%%%%%%% mcdc_symbols 

%% dc line parameters and vairables
\newcommand{\Pdcdef}{P^{\text{\dcs}}_{d^{\phi}e^{\phi}f^{\phi}}}
\newcommand{\Pdcdefo}{P^{\text{\dcs}}_{d^{0}e^{0}f^{0}}}
\newcommand{\Pdcdfe}{P^{\text{\dcs}}_{d^{\phi}f^{\phi}e^{\phi}}}
\newcommand{\Pdclossd}{P^{\text{\dcs,loss}}_{d^{\phi}}}
\newcommand{\Pdcratedd}{P^{\text{\dcs,rated}}_{d^{\phi}}}
\newcommand{\Idcdef}{I^{\text{\dcs}}_{d^{\phi}e^{\phi}f^{\phi}}}
\newcommand{\Idcdfe}{I^{\text{\dcs}}_{d^{\phi}f^{\phi}e^{\phi}}}
\newcommand{\Idcdrated}{I^{\text{\dcs,rated}}_{d^{\phi}}}
\newcommand{\Idcdmax}{I^{\text{\dcs,max}}_{d^{\phi}}}
\newcommand{\Idcdmin}{I^{\text{\dcs,min}}_{d^{\phi}}}
\newcommand{\idcdef}{i^{\text{sq,\dcs}}_{d^{\phi}e^{\phi}f^{\phi}}}
\newcommand{\Ue}{U^{\text{\dcs}}_{e^{\phi}}}
\newcommand{\Uemin}{U^{\text{\dcs,max}}_{e^{\phi}}}
\newcommand{\Uemax}{U^{\text{\dcs,max}}_{e^{\phi}}}
\newcommand{\Uf}{U^{\text{\dcs}}_{f^{\phi}}}
\newcommand{\gd}{g_{d^{\phi}}}
\newcommand{\rd}{r_{d^{\phi}}}
\newcommand{\pd}{p_{d^{\phi}}}
\newcommand{\rdo}{r_{d0}}
\newcommand{\rg}{r_{g}}
\newcommand{\Ueo}{U^{\text{\dcs}}_{e^{0}}}

%
%
% converter parameters and variables
\newcommand{\Pconvac}{P^{\text{\convs,\acs}}_{c^{\rho}}}
\newcommand{\Pconvacp}{P^{\text{\convs,\acs}}_{c^{1}}}
\newcommand{\Pconvacn}{P^{\text{\convs,\acs}}_{c^{2}}}
\newcommand{\Pconvacmin}{P^{\text{\convs,\acs,min}}_{c^{\rho}}}
\newcommand{\Pconvacmax}{P^{\text{\convs,\acs,max}}_{c^{\rho}}}
\newcommand{\Qconvac}{Q^{\text{\convs,\acs}}_{c^{\rho}}}
\newcommand{\Qconvacmin}{Q^{\text{\convs,\acs,min}}_{c^{\rho}}}
\newcommand{\Qconvacmax}{Q^{\text{\convs,\acs,max}}_{c^{\rho}}}
\newcommand{\Sconvacrated}{S^{\text{\convs,\acs,rated}}_{c^{\rho}}}
\newcommand{\Pconvdc}{P^{\text{\convs,\dcs}}_{c^{\rho}}}
\newcommand{\Pconvdcp}{P^{\text{\convs,\dcs}}_{c^{1}}}
\newcommand{\Pconvdcpo}{P^{\text{\convs,\dcs}}_{c^{10}}}
\newcommand{\Pconvdcn}{P^{\text{\convs,\dcs}}_{c^{2}}}
\newcommand{\Pconvdcno}{P^{\text{\convs,\dcs}}_{c^{20}}}
\newcommand{\Pconvdco}{P^{\text{\convs,\dcs}}_{c^{\rho0}}}
\newcommand{\Pconvdcmin}{P^{\text{\convs,\dcs,min}}_{c^{\rho}}}
\newcommand{\Pconvdcmax}{P^{\text{\convs,\dcs,max}}_{c^{\rho}}}
\newcommand{\Pconvloss}{P^{\text{\convs,loss}}_{c^{\rho}}}
\newcommand{\Pconvlossp}{P^{\text{\convs,loss}}_{c^{1}}}
\newcommand{\Pconvlossn}{P^{\text{\convs,loss}}_{c^{2}}}
\newcommand{\Iconvmag}{I^{\text{\convs,mag}}_{c^{\rho}}}
\newcommand{\Iconvmaglin}{I^{\text{lin,\convs,mag}}_{c^{\rho}}}
\newcommand{\Iconvmagsq}{i^{\text{sq,\convs,mag}}_{c^{\rho}}}
\newcommand{\Iconvrated}{I^{\text{\convs,rated}}_{c^{\rho}}}
\newcommand{\aconv}{a^{\text{\convs}}_{c^{\rho}}}
\newcommand{\bconv}{b^{\text{\convs}}_{c^{\rho}}}
\newcommand{\cconv}{c^{\text{\convs}}_{c^{\rho}}}

\newcommand{\Iconvdcmag}{I^{\text{\convs,dc,mag}}_{c^{\rho}}}
\newcommand{\Iconvdcmin}{I^{\text{\convs,dc,min}}_{c^{\rho}}}
\newcommand{\Iconvdcmax}{I^{\text{\convs,dc,max}}_{c^{\rho}}}

\newcommand{\Iconvdc}{I^{\text{\convs,\dcs}}_{c^{\rho}}}
\newcommand{\Iconvdcp}{I^{\text{\convs,\dcs}}_{c^{1}}}
\newcommand{\Iconvdcpo}{I^{\text{\convs,\dcs}}_{c^{10}}}
\newcommand{\Iconvdcn}{I^{\text{\convs,\dcs}}_{c^{2}}}
\newcommand{\Iconvdcno}{I^{\text{\convs,\dcs}}_{c^{20}}}
\newcommand{\Iconvdco}{I^{\text{\convs,\dcs}}_{c^{0}}}
\newcommand{\Iconvdcg}{I^{\text{\convs,\dcs}}_{c_{g}}}

\newcommand{\Ufilti}{U^{\text{\filter}}_{c^{\rho}}}
\newcommand{\Ufiltmagi}{U^{\text{\filter,mag}}_{c^{\rho}}}
\newcommand{\thetafilti}{\theta^{\text{\filter}}_{c^{\rho}}}

\newcommand{\Uconvi}{U^{\text{\convs}}_{c^{\rho}}}
\newcommand{\Uconvmagi}{U^{\text{\convs,mag}}_{c^{\rho}}}
\newcommand{\thetaconvi}{\theta^{\text{\convs}}_{c^{\rho}}}

% loads and generators
\newcommand{\Pl}{P_{m}}
\newcommand{\Ql}{Q_{m}}
\newcommand{\Bi}{b^{\text{shunt}}_{i}}
\newcommand{\Gi}{g^{\text{shunt}}_{i}}

%%transformer, filter, reactor

\newcommand{\Ztf}{z^{\text{\transfs}}_{c^{\rho}}}
\newcommand{\Rtf}{r^{\text{\transfs}}_{c^{\rho}}}
\newcommand{\Xtf}{x^{\text{\transfs}}_{c^{\rho}}}
\newcommand{\Ytf}{y^{\text{\transfs}}_{c^{\rho}}}
\newcommand{\Gtf}{g^{\text{\transfs}}_{c^{\rho}}}
\newcommand{\Btf}{b^{\text{\transfs}}_{c^{\rho}}}

\newcommand{\Ptf}{P^{\text{\transfs}}_{c^{\rho}ie^{\rho}}}
\newcommand{\Qtf}{Q^{\text{\transfs}}_{c^{\rho}ie^{\rho}}}
\newcommand{\Ptfei}{P^{\text{\transfs}}_{c^{\rho}e^{\rho}i}}
\newcommand{\Qtfei}{Q^{\text{\transfs}}_{c^{\rho}e^{\rho}i}}
\newcommand{\Itf}{I^{\text{\transfs}}_{c^{\rho}}}
\newcommand{\Ptfp}{P^{\text{\transfs}}_{c^{1}ie^{1}}}
\newcommand{\Ptfn}{P^{\text{\transfs}}_{c^{2}ie^{2}}}
\newcommand{\Qtfp}{Q^{\text{\transfs}}_{c^{1}ie^{1}}}
\newcommand{\Qtfn}{Q^{\text{\transfs}}_{c^{2}ie^{2}}}

\newcommand{\Zpr}{z^{\text{\reactor}}_{c^{\rho}}}
\newcommand{\Rpr}{r^{\text{\reactor}}_{c^{\rho}}}
\newcommand{\Xpr}{x^{\text{\reactor}}_{c^{\rho}}}
\newcommand{\Ypr}{y^{\text{\reactor}}_{c^{\rho}}}
\newcommand{\Gpr}{g^{\text{\reactor}}_{c^{\rho}}}
\newcommand{\Bpr}{b^{\text{\reactor}}_{c^{\rho}}}
\newcommand{\Ppr}{P^{\text{\reactor}}_{c^{\rho}ie^{\rho}}}
\newcommand{\Qpr}{Q^{\text{\reactor}}_{c^{\rho}ie^{\rho}}}
\newcommand{\Pprei}{P^{\text{\reactor}}_{c^{\rho}ie^{\rho}}}
\newcommand{\Qprei}{Q^{\text{\reactor}}_{c^{\rho}ie^{\rho}}}
\newcommand{\Ipr}{I^{\text{\reactor}}_{c^{\rho}}}
\newcommand{\Bf}{b^{\text{\filter}}_{c^{\rho}}}
\newcommand{\Qf}{Q^{\text{\filter}}_{c^{\rho}}}
\newcommand{\tc}{t_{c^{\rho}}}

%% file: full_paper.tex
\section{Introduction} \label{section: introduction}
The monopolar operation of a point-to-point bipolar HVDC link is well explored and utilized \cite{NordLink_whitepaper}; however, the same for multi-terminal (MTDC) and meshed bipolar systems, also referred to as HVDC grids, remains less explored. Nordic link, the first bipolar Voltage Source Converter (VSC) High Voltage Direct Current (HVDC) system at $\pm~525~kV$ voltage level, is a rigid bipole system without dedicated metallic return \cite{NordLink_whitepaper}. Although this design scheme ensures half of the rated capacity in case of a single converter outage, it does not enable monopolar operation (without using ground return) during single-line conductor outages. Therefore, realizing the value of these monopolar operations of the bipolar systems during contingencies, numerous recent projects are configured with a dedicated metallic return, e.g., Tennet's 2~GW program \cite{Tennet2GW}.

The early stage of the HVDC grid concept evolved with either monopolar or balanced bipolar configurations \cite{VanHertem2016, wang2020multi}. Therefore, the power flow problem for hybrid AC/DC systems with a balanced DC side has been explored extensively in the last decade \cite{beerten2012generalized, 6702505, 6583292, 6588621, 9546672}. Beerten et al. \cite{beerten2012generalized} present a steady-state model for power flow analysis in VSC-based MTDC systems. The power flow model in \cite{beerten2012generalized} considers MTDC systems with strictly one converter controlling the DC voltage, thus acting as DC slack, while the other converters operate in constant DC power mode. The model is further extended into distributed voltage slack using DC-voltage droop control mode at the individual converters, as presented in \cite{6702505}, \cite{6583292} and \cite{6588621}.  Fernández-Pérez et al. present a linear power flow model also incorporating the system losses in VSC MTDC systems \cite{9546672}. A modified AC/DC power flow algorithm with improved convergence is discussed in \cite{7961222}. However, all the above models and algorithms assume a single-conductor representation of the DC grid and converter stations, making them applicable only to monopolar configurations or balanced bipolar HVDC systems.
 
In a monopolar DC grid, an outage of a converter pole or DC branch conductor does not cause an unbalanced operation. Contrary to that, an asymmetric monopolar operation of a DC branch or converter in the meshed or multi-terminal bipolar system results in unbalanced voltages and power flows in the DC network. Therefore, an AC/DC power flow model for bipolar HVDC grids is needed for power flow analysis during normal modes of operation as well as during contingencies.
 
Lee et al. present a power flow model for bipolar DC grids \cite{lee2022generic}, however, the work is focused on purely DC grids and, therefore, does not model the AC/DC converters and the AC grid. Li et al. \cite{li2021research} present a power flow calculation method for bipolar HVDC systems but focuses solely on DC-side control modes and excludes a detailed representation of AC/DC converter stations. Early work by the authors \cite{10304389} introduces a multi-conductor model with an explicit representation of each converter pole and conductor of the DC branches, including the metallic or ground return path. Thus, this model accurately represents the balanced as well as unbalanced bipolar DC grids (comprising mixed HVDC monopolar and bipolar configurations). However, the model in \cite{10304389} does not incorporate the control modes of AC/DC converters since it focuses on finding an optimal operating point for the unbalanced system while respecting the component limits. 

\begin{figure}
        \centering
        \includegraphics[width=0.65\columnwidth]{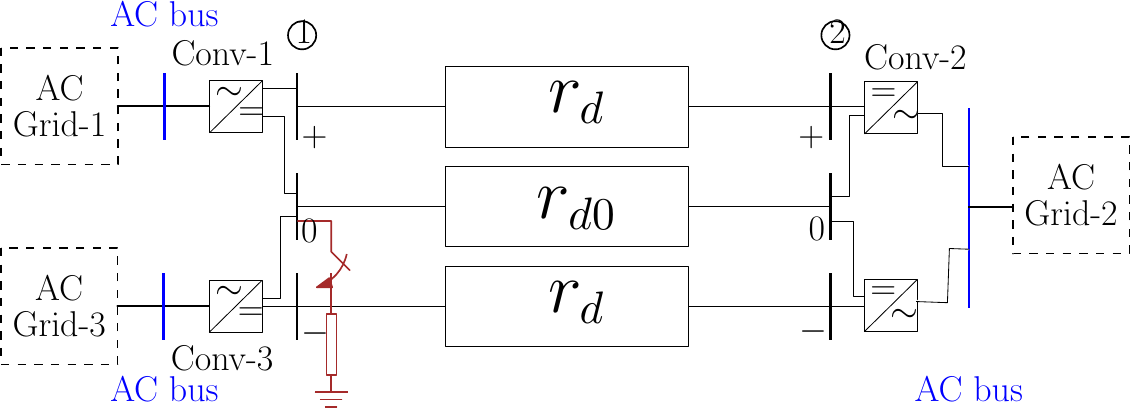}
        \caption{A bipolar HVDC link connecting three asynchronous AC systems}
        \label{fig:test_case} \vspace{-0.2cm}
    \end{figure}

As the real-world power system conditions often deviate from those assumed during optimization, the actual operating point is primarily governed by system controllers, including converter controls. Therefore, it is important to model converter control modes in an AC/DC power flow model. The basic control strategies for the DC side of an AC/DC converter are usually classified as constant DC voltage control, constant active power control, and DC voltage-power droop control \cite{vrana2013classification}. The impact of DC-droop control on power sharing and network voltages is discussed in \cite{6583292}, particularly for converter contingencies. Numerous other works explored how to choose the droop coefficients used in the droop control \cite{lee_method_2020, zhang_minimization_2021, tavakoli_dc_2020}. As a VSC-based converter can independently control active and reactive powers, the AC side of the converter can also be operated in three similar control modes, i.e., constant AC voltage magnitude control, constant reactive power control, and AC voltage-reactive power droop. The droop control on the AC side is important in the context of bipolar HVDC grids, where reactive power support to the AC grid can be shared among the two converters connected to the same AC bus as for Conv-2 in \cref{fig:test_case}. Although there exist more advanced control modes in the literature, most of these control strategies can be represented with the previously mentioned control modes \cite{vrana2013classification}. This paper aims to incorporate the basic control strategies into a power flow problem for bipolar HVDC grids in hybrid AC/DC systems.
 
The power flow problem for hybrid AC/DC systems is generally modeled with either a unified or a sequential approach. In the unified approach, all the equations related to the AC network, DC network, and AC/DC converter are grouped to form a set of non-linear algebraic equations. Meanwhile, in the sequential approach, the AC and DC networks are solved separately in each iteration, where the output of one acts as fixed input to the other, along with convergence checks. While the sequential methods are easy to integrate into the existing algorithms, they have poor convergence compared to the unified methods \cite{7961222}. Therefore, some researchers opt for unified methods \cite{baradar2013multi, 6189804, lei2016general, 9268999} while some rely upon sequential methods with advanced techniques to improve convergence \cite{7961222, beerten2012generalized}. In this paper, an optimization-based approach is used, considering all the system equations simultaneously, i.e., a unified approach.   

The major contributions of this paper are as follows. 
\begin{itemize}
    \item A unified power flow model for the unbalanced operation of bipolar HVDC grids is presented, enabling power flow analysis and contingency studies with unbalanced HVDC sides in AC/DC systems. The model incorporates all basic control modes of VSCs in HVDC systems. 
    \item The concept of DC-droop control, defined as the droop relationship between DC voltage and converter active power, is extended to bipolar HVDC grids. 
    
    \item The proposed model incorporates an AC-droop control mode, representing the droop relationship between AC voltage magnitude and reactive power for bipolar HVDC grids. This control mode facilitates desired reactive power sharing between two or more converters connected to the same AC bus.
    
    \item The power flow model is implemented as an open-source tool (in the Julia/JuMP framework).
\end{itemize}

Section II defines the proposed power flow model with the detailed modeling of various converter control modes. Section III presents the test case and numerical results, including an analysis of different control modes during normal operation and a converter contingency. Finally, Section IV states the conclusions. 

\section{Power Flow Model for Bipolar AC/DC grids }
The power flow (or load flow) problem in power systems involves calculating nodal voltages (magnitude and angle in polar form) and, thus, network flows, given knowledge of network configuration, network parameters, and nodal injections. This problem is mathematically represented as a set of non-linear algebraic equations, comprising network flow equations and control setpoints for node voltages or injections. The following subsections present a detailed formulation of the power flow equations for hybrid AC/DC systems. 

AC nodes are represented by \(i, j \in \mathcal{I}\) and AC branches by \(l \in \mathcal{L}\), with AC topology defined by \( \TopoAC \subseteq \mathcal{L} \times \mathcal{I} \times \mathcal{I} \). DC nodes are denoted by \(e, f \in \mathcal{E}\), and DC branches by \(d \in \mathcal{D}\), with DC topology given by \( \TopoDC \subseteq \mathcal{D} \times \mathcal{E} \times \mathcal{E} \). AC/DC converters are indexed by \(c \in \mathcal{C}\) and their poles by \( \rho \in \{1,2\} \), with the converter topology defined by \( \Topoconv \subseteq \mathcal{C} \times \mathcal{I} \times \mathcal{E} \). Generators, reference AC buses, and loads are indexed by \(g \in \mathcal{G}\), \(r \in \mathcal{R}\), and \(m \in \mathcal{M}\), respectively. The connectivity of AC generators, AC loads, and DC loads to their respective nodes is represented by \( \mathcal{T}^{\text{gen,\acs}} \subseteq \mathcal{G} \times \mathcal{I} \), \( \mathcal{T}^{\text{load,\acs}} \subseteq \mathcal{M} \times \mathcal{I} \), and \( \mathcal{T}^{\text{load,\dcs}} \subseteq \mathcal{M} \times \mathcal{E} \).

\subsection{Network Flow Equations}
\begin{figure}
    \centering
    \includegraphics[width=\textwidth]{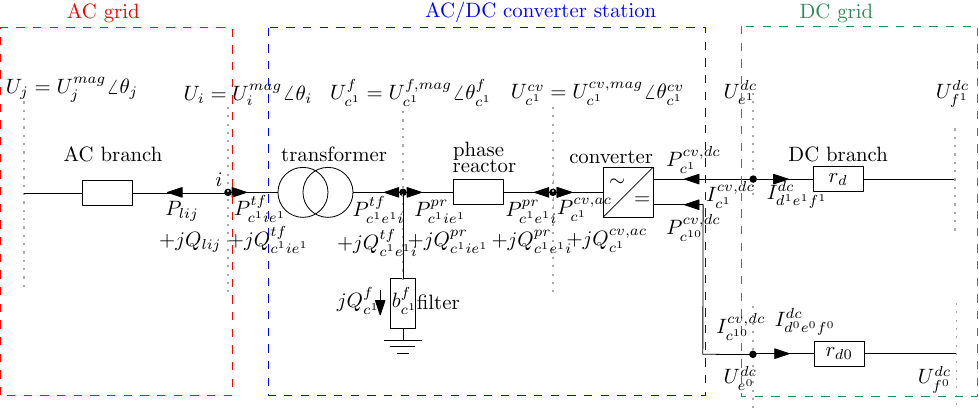}
    \caption{An AC/DC converter station connected between the positive and neutral terminals on the DC side }
    \label{fig: ac_dc_bipolar}
\end{figure}

For a hybrid AC/DC system, the branch flow and nodal power balance equations can be divided into three sets as shown in Fig. \ref{fig: ac_dc_bipolar}, where a converter is connected between positive and neutral terminals on the DC side. A bipolar converter station (Conv-2 in Fig. \ref{fig:test_case}) can also be represented similarly. 
  
    \subsubsection{AC grid}
    The active and reactive power flow in the AC branch $l$, from node $i$ to $j$ are
    \begin{align}
         \Plij =  g_{lij}\cdot({\Umagi}^2 - {\Umagi} {\Umagj}\cdot \cos(\thetai-\thetaj)) \nonumber \\
        \quad -b_{lij}\cdot({\Umagi}\cdot{\Umagj}\cdot \sin(\thetai-\thetaj)) \quad\quad \forall lij \in \TopoAC \label{eq:plij_acp}\\
         \Qlij = -b_{lij}\cdot({\Umagi}^2 - {\Umagi}\cdot{\Umagj}\cdot \cos(\thetai-\thetaj)) \nonumber \\
         \quad-g_{lij}\cdot({\Umagi}\cdot{\Umagj}\cdot \sin(\thetai-\thetaj)) \quad\quad \forall lij \in \TopoAC
         \end{align}
    where $y_{lij}= g_{lij} + j\cdot b_{lij}$ represents the branch admittance.  
        Kirchhoff's Current Law in terms of active and reactive power balance on AC grid nodes $i \in \TopoAC$ is defined as \vspace{-0.3cm}
    \begin{align}
       \quad\quad \sum_{cie \in \Topoconv, \rho \in \{1,2\}} \Ptf + \quad\quad \!\!\!\sum_{lij \in \TopoAC}  \Plij  \nonumber \\
        = \!\! \sum_{gi \in\TopogenAC} \!\!\Pg -\!\!\!\sum_{mi \in \TopoloadAC } \!\!\Pl - \Gi (\Umagi)^2~~~\forall i \in \mathcal{I} \label{eq:kcl_acp},
    \end{align} \vspace{-0.6cm}
    \begin{align}	
    \quad\quad \sum_{cie \in \Topoconv, \rho \in \{1,2\}} \Qtf + \sum_{lij \in \TopoAC}  \Qlij \nonumber \\
        =\!\! \sum_{gi \in \TopogenAC} \!\!\Qg -\!\!\!\sum_{mi \in  \TopoloadAC } \!\!\Ql +\Bi (\Umagi)^2~~~\forall i \in \mathcal{I} \label{eq:kcl_acq}
    \end{align}
    where $\Pg$, $\Pl$, $\Qg$, and $\Ql$ represent the active and reactive power setpoints of generators and loads respectively. Similarly, $\Ptf$ and $\Qtf$ are the active and reactive powers for pole $\rho$ of converter $c$ connected to AC bus $i$. The variable $\Gi+j\Bi$ is the shunt admittance connected to the AC bus.

    \subsubsection{DC grid}
    This paper models the DC grid in a current-voltage variable, i.e., I-V formulation for numerical robustness in a system with near zero voltages (at neutral terminals), as discussed in \cite{10304389}. The nodal current balance for the positive and negative terminals of a DC bus $e$ is   \vspace{-0.3cm}
        \begin{align} 
        \!\!\!\! \sum_{cie \in \Topoconv}\!\!\!\!  \Iconvdc 
        + \sum_{def \in \TopoDC}\!\!\!\!  \Idcdef = 0 ~~\forall e \in \mathcal{E}, \phi \in {\{1,2\}} \label{eq:kcl_dc_c},
        \end{align} 
        whereas the additional converter grounding is also modeled for the neutral terminal (i.e. $\phi = 0$) as \vspace{-0.3cm}
        \begin{align}
        \!\!\!\! \sum_{cie \in \Topoconv} I^{\text{\convs,\dcs}}_{c^{\rho0}} + \!\!\!\! \sum_{cie \in \Topoconv} \Iconvdcg +\!\!\!\! \sum_{def \in \TopoDC} \Idcdef = 0 ~~\forall e \in \mathcal{E} \label{eq:kcl_dc_cg}.
        \end{align}
    The DC branch currents in (\ref{eq:kcl_dc_c}) and (\ref{eq:kcl_dc_cg}) are given as \vspace{-0.3cm}
        \begin{align}
        \Idcdef = (1/\rd) \cdot (\Ue - \Uf )~~\forall def \in \TopoDC \cup \TopoDCrev, ~~~~~  \label{eq:BIM}
        \end{align}
        for all three conductors (i.e. $\phi \in {\{1,2\}}$). The converter currents in (\ref{eq:kcl_dc_c}) and (\ref{eq:kcl_dc_cg}) are defined in the next section.  
        
    \subsubsection{AC/DC converter}
        Nodal power balances and branch flow equations for an AC/DC converter are defined in this section. At the Point of Common Coupling (PCC), the converter transformer of pole $\rho$ of converter $c$ has the active and reactive power defined as  
        \begin{align}
        	\Ptf = \Gtf \left(\frac{\Umagi}{\tc} \right)^2  -\Gtf \frac{\Umagi}{\tc} \Ufiltmagi \cos(\thetai-\thetafilti)   \nonumber \\
        	-\Btf \frac{\Umagi}{\tc} \Ufiltmagi \sin(\thetai-\thetafilti)~~\forall  cie \in \Topoconv , \label{tr_start_p}\\
        	\Qtf =-\Btf \left(\frac{\Umagi}{\tc}\right)^2 +  \Btf \frac{\Umagi}{\tc} \Ufiltmagi \cos(\thetai- \thetafilti)    \nonumber \\
        	-\Gtf \frac{\Umagi}{\tc} \Ufiltmagi \sin(\thetai-\thetafilti)~~\forall  cie \in \Topoconv, ~~~~ \label{tr_start_q}
        \end{align}
        seen at the AC bus $i$. Here $\Gtf$ and $\Btf$ are the conductance and susceptance of the converter transformers for the $\rho^{th}$ pole, respectively. Variable $\tc$ is the converter transformer tap changer setting. Similar equations are also defined for the reverse direction at the filter bus connection point of the transformer (i.e.  $e \rightarrow i$) to calculate $\Ptfei$ and $\Qtfei$.
        
        The reactive power of the filter capacitor of pole  $\rho$ of the converter $c$ is calculated as  
        \begin{align}
        	\Qf = - \Bf (\Ufiltmagi)^2~~~~\forall  cie \in \Topoconv. 
        \end{align}
        
    The phase reactor impedance is represented as $\Zpr = \Rpr + j \Xpr$, and its equivalent admittance is given by $\Ypr = \frac{1}{\Zpr} = \Gpr + j \Bpr$. Since both the transformer and the reactor are modeled as a series impedance, equations (\ref{tr_start_p}) - (\ref{tr_start_q}) can be employed to model the phase reactor by setting $\tc = 1$ and using the corresponding voltage variables and angles.
   
        The active and reactive power balance at the node connecting the transformer, the filter capacitor, and the phase reactor is defined as
        \begin{align}
        	\Ppr +  \Ptfei= 0  ~~\forall  cie \in \Topoconv, \label{eq_p_filter_balance} \\
        	\Qpr +  \Qtfei + \Qf= 0~~\forall  cie \in \Topoconv.  \label{eq_q_filter_balance}
        \end{align}
        
        The AC and DC side power injections of the converter poles are linked through the converter losses:
        \begin{align}
        \Pconvac + \Pconvdc + \Pconvdco = \Pconvloss  ~~ \forall  c \in \mathcal{C}. \label{eq:conv_ac_dc1}
        \end{align}
       where $\Pconvac \text{and} \Pconvdc$ denotes the power flows into the converter as seen in Fig. \ref{fig: ac_dc_bipolar}.  $\Pconvloss$ is the converter power loss defined as
        \begin{align}
        \!\Pconvloss = \aconv + \bconv \cdot \Iconvmag + \cconv \cdot (\Iconvmag)^2~\forall  c \in \mathcal{C} \label{eq:conv_losses}
        \end{align}
        for both positive and negative poles. $\Iconvmag$ is the magnitude of converter current calculated as
        \begin{align}
         (\Pconvac)^2 + (\Qconvac)^2  \!= \ (\Uconvmagi)^2 \cdot (\Iconvmag)^2~\forall  c \in \mathcal{C} \label{eq:conv_loss_ac},
        \end{align}

        $\Pconvdc$ and $\Pconvdco$ in (\ref{eq:conv_ac_dc1}) are linked to the current and voltage variable on the DC side as
        \begin{align}
        \Pconvdc = \Ue \cdot \Iconvdc ~~~~\forall  c \in \mathcal{C}, \label{eq:conv_ac_d_dc}\\
        \Pconvdco = \Ueo \cdot I^{\text{\convs,\dcs}}_{c^{\rho0}} ~~~~\forall  c \in \mathcal{C}, \label{eq:conv_ac_d_dcg_lim}
        \end{align}
        Whereas DC currents of a bipolar configuration satisfy the following relation
        \begin{align}
        \Iconvdcp + \Iconvdcpo = 0 \label{eq_Iconvdc_p},\\
        \Iconvdcn + \Iconvdcno = 0 \label{eq_Iconvdc_n},\\
        \Iconvdco = \Iconvdcpo + \Iconvdcno \label{eq_Iconvdc_o},
        \end{align}
        for converter $c$, the superscripts $1$ and $2$ refer to the positive and negative poles, respectively, while $0$ denotes the neutral.
        
    \subsection{Node voltage and injection setpoints} \label{sec: control_setpoints}
    Mathematically the set of algebraic equations from (\ref{eq:plij_acp}) to (\ref{eq_Iconvdc_o}) is under-determined, i.e., the number of variables (unknowns) is more than the number of equations, and therefore some of the variables need to be specified \cite{grainger1999power}. From the power system point of view, this is achieved by specifying some of the nodal voltages and injection setpoints for controllable quantities for a desired operating state of the system. 
    \subsubsection{Generator and load setpoints}
    In the power flow problem, generator buses are modeled as constant active power injection while maintaining the bus voltage magnitude at the desired value. Thus, generator-connected buses are typically classified as PV buses, except for the reference bus, which serves as the system's voltage angle reference \cite{grainger1999power}. The loads are modeled as constant active and reactive power withdrawal (negative injection) i.e. both magnitude and angle of voltage remain variables at load buses without any generator or voltage control device. The already known or specified quantities for load and generator bus are mathematically added to the model as \vspace{-0.3cm}
        \begin{align}
        	\theta_{ref} = 0, \\
             P_{g}, U^{mag}_{j,gen} = constant, \\
              P_{m}, Q_{m} = constant. 
        \end{align}
    \subsubsection{AC/DC converter control modes} \label{sec: control_modes}

       \begin{figure}[t!]
        \centering
        \includegraphics[]{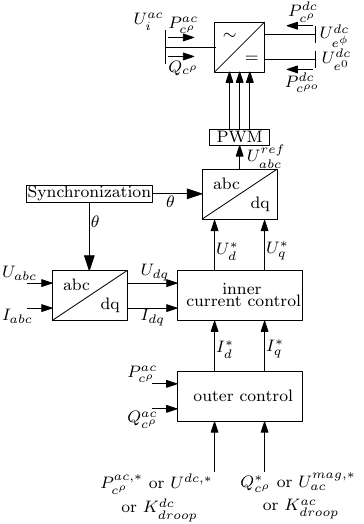}
        \caption{VSC-HVDC converter control architecture}
        \label{fig: conv_control_concept}
    \end{figure}
                     
        \begin{figure}[t!]
        \centering
        \includegraphics[]{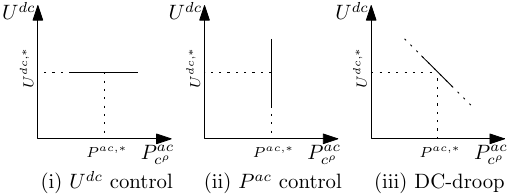}
        \caption{$d$-axis control modes of a VSC-converter}
        \label{fig: control_diagrams_dc}
     \end{figure}

     \begin{figure}[t!]
        \centering
        \includegraphics[]{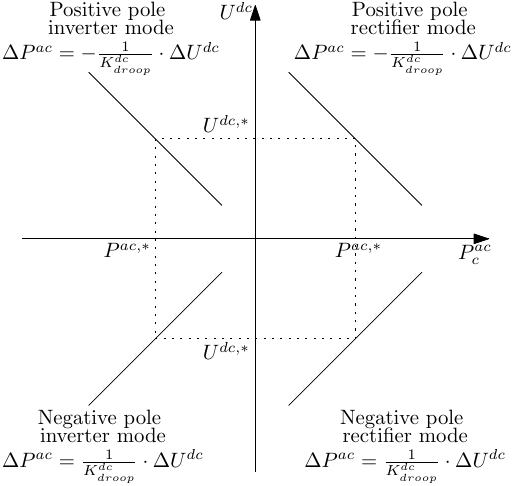}
        \caption{$d$-axis DC-droop control mode in bipolar HVDC systems}
        \label{fig: control_diagrams_dc_droop_bipolar}
     \end{figure}

             \begin{figure}[t!]
        \centering
        \includegraphics[]{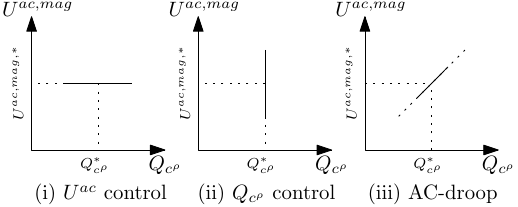}
        \caption{$q$-axis control modes of a VSC-converter}
        \label{fig: control_diagrams_ac}
     \end{figure}        
        In hybrid AC/DC power systems, the converter acts as an additional control element in the system. The current flowing from the AC terminals to a VSC HVDC converter is typically controlled in a decoupled $dq$ frame as shown in \cref{fig: conv_control_concept}. A VSC can control active and reactive power independently within the converter's apparent power limits. In this paper, the active power ($P^{ac}$) or DC bus voltage ($U^{dc}$) controls provide the $d$-axis current reference. The $q$-axis current reference is determined by the reactive power ($Q$) or the AC side bus voltage magnitude ($U^{ac,mag}$) controls. For a given operating (control) mode of the converter, its control action must be defined for both the $d$-axis and $q$-axis. There can be various ways to design the converter control for different objectives. In this paper, three basic control techniques of outer control are incorporated into the steady-state power model for unbalanced bipolar HVDC grids. In this model, the active power $P^{ac}$ and the reactive power $Q$ are controlled at the PCC, and therefore, they correspond to $\Ptf$ and $\Qtf$ in network flow equations \eqref{tr_start_p} and \eqref{tr_start_q} respectively. 
        
       \subsubsection*{$d$-axis control}
        
        The three basic control modes for the $d$-axis are DC voltage control ($U^{dc}$), active power control (${P^{ac}}$), and $U^{dc}-{P^{ac}}$ droop (DC-droop), as shown in \cref{fig: control_diagrams_dc}. The choice of a control mode in a bipolar system translates into a mathematical equation as follows:         
                \begin{equation} 
                    U^{dc}_{e^\phi}-U^{dc}_{e^{0}} = U^{dc,*}_{e^\phi},  ~~~~~~ \phi \in {\{1,2\}} \label{eq:vdc_ref}
                    \end{equation} 
                    \begin{equation} 
                    P^{ac}_{c^\rho} =  P^{ac,*}_{c^\rho}, ~~~~~~ \rho \in {\{1,2\} }\label{eq:Pdc_ref}
                \end{equation} 
                where the reference voltage of terminal $\phi$ of DC bus $e$, connected to pole $\rho$ of converter $c$, is defined with respect to its neutral terminal ($\phi = 0$). The active power is controlled on the AC side at the PCC, with direction as indicated in \cref{fig: conv_control_concept}. The mathematical expression of (\ref{eq:vdc_ref}) and (\ref{eq:Pdc_ref}) stays the same irrespective of the DC terminal polarity (positive or negative) and converter operating mode (rectifier or inverter). However, the DC-droop equation must be adjusted based on the voltage polarity of the DC bus terminal to which the converter (pole) is connected. This adjustment is not influenced by the converter's operating mode, whether rectifier or inverter. The mathematical equation for a positive terminal-connected converter is
                \begin{align} 
                    P^{ac}_{c^\rho} - P^{ac,*}_{c^\rho} =   -  \frac{1}{K^{dc}_{droop,{c^\rho}}} \cdot ((U^{dc}_{e^\rho}-U^{dc}_{e^{0}}) -U^{dc,*}_{e^\rho}),
                    \label{eq:dc_droop_p}
                \end{align}
                while that for the negative terminal-connected converter is
                \begin{align} 
                    P^{ac}_{c^\rho} - P^{ac,*}_{c^\rho} =  \frac{1}{K^{dc}_{droop,{c^\rho}}} \cdot ((U^{dc}_{e^\rho}-U^{dc}_{e^{0}}) -U^{dc,*}_{e^\rho})
                    \label{eq:dc_droop_n}.
                \end{align}
                A graphical representation of the same is depicted in \cref{fig: control_diagrams_dc_droop_bipolar}. 
    
        \subsubsection*{$q$-axis control}
     
        Similar to $d$-axis controls, the $q$-axis can also have three basic control modes i.e., AC voltage ($U^{ac,mag}$) control, reactive power ($Q$) control, and $U^{ac,mag}-Q$ droop (AC-droop) as depicted in \cref{fig: control_diagrams_ac}. The mathematical equations are written as 
                \begin{equation}  
                \Umagi = U_{i}^{mag,*}  \label{eq:vac_ref}
                    \end{equation} 
                                \begin{equation} 
                        Q_{c^{\rho}} = Q^{*}_{c^{\rho}} \label{eq:Qdc_ref}\end{equation} 
                                \begin{equation} 
                \!\!  Q_{c^{\rho}}- Q^{*}_{c^{\rho}} = \frac{1}{K^{ac}_{droop,{c^\rho}}} \cdot  (\Umagi-U_{i}^{mag,*}) \label{eq:ac_droop}
                \end{equation} 
        Here, (\ref{eq:ac_droop}) represents the AC-droop control mode for bipolar HVDC grids which has not been explored in the existing literature. This is particularly relevant in the context of unbalanced operation where two converters connected to the same AC bus can share the reactive power support. The slope of (\ref{eq:ac_droop}) is positive for the sign convention as shown in \cref{fig: conv_control_concept} i.e., if the magnitude of the AC bus voltage increases beyond its reference value, the converter reactive power (withdrawn at the bus) increases (to limit the rise). This equation stays the same for all the converters irrespective of their operating mode and DC side connection.

    \subsection{Power flow in bipolar HVDC and AC-droop control mode} \label{sec: AC-droop}
        
        Bipolar DC grids have three voltage polarities i.e., positive, negative, and neutral. It can also be seen as two voltage layers (or poles, used interchangeably) with a common neutral. In a power flow model, at least one of the bus terminals for each of these three polarities should be fixed to a predefined voltage level to provide the references. The converter station grounding provides this reference for the neutral polarity as 
                  \begin{align}  
                      \Ueo = 0 \ \label{eq:dc_neutral_ground}
                    \end{align}
        \cref{eq:dc_neutral_ground} is applied for the neutral terminal of the DC bus $e$ which is grounded. For positive and negative polarities, the reference is provided by the converter operating in voltage control or DC slack mode. Thus, it is essential to have at least one converter operating to control DC voltage for each layer (positive and negative). In the case of a single DC slack in each layer, it is usually provided by DC voltage control (\ref{eq:vdc_ref}), whereas in the case of distributed slack, this is provided by DC-droop control (\ref{eq:dc_droop_p}) or (\ref{eq:dc_droop_n}).
    
        This voltage reference is essential for both balanced and unbalanced operation of bipolar DC systems. However, it is not so explicitly visible in single-conductor models for bipolar HVDC grids, because such models assume both (positive and negative pole) converters of bipolar converter stations to operate in the same control mode. Further, the neutral terminal is not modeled and is assumed to be at zero voltage in single-conductor models. The assumption of the same control modes also fails to distinguish infeasible control modes. For example, two converters connected to the same AC bus (Conv-2 in \cref{fig:test_case}) might lead to unstable and impractical results if operated in the AC-voltage control mode at the same time. Since the single conductor DC model-based power flow tools do not explicitly consider both poles, they would find a solution for such systems, which might not be feasible (when implemented with integral controllers). Conversely, the multiconductor DC model-based solution, for such a scenario, would not be unique because the amount of the reactive power shared by each converter would not be unique in the absence of a reactive power-sharing mechanism. In such a scenario, operating one converter pole in AC-voltage control mode and another in reactive power control mode can be an option. Alternatively, introducing a droop relation between AC bus voltage magnitude and converter reactive power (as shown in (\ref{eq:ac_droop}) and \cref{fig: control_diagrams_ac} (iii)) also enables the reactive power sharing among the converters connected to the same AC bus. 
        
    \subsection{Unified power flow methodology}
        The set of non-linear algebraic equations, i.e., power flow problem formulated by (\ref{eq:plij_acp}) to (\ref{eq:dc_neutral_ground}) is solved with an optimization-based approach using Ipopt v3.1.1.4 \cite{wachter2006implementation}, which is an open-source non-linear optimization solver. As demonstrated in \cite{8442948}, an optimization problem can be reduced to a root-finding problem by adding the appropriate constraints, thus shrinking the feasible space from a continuous region to discrete points. In a standard unconstrained optimization problem as 
        \begin{align}
                &\min _{\mathbf{x}} f(\mathbf{x}),
                % \end{align}
                  \hspace{15mm} \textit{subject to} 
                % \begin{align}
                &g(\mathrm{x, y})=0.
               \end{align}
        If the equality constraint $g(\mathrm{x, y})=0$ consists of all the algebraic equations from (\ref{eq:plij_acp}) to (\ref{eq:dc_neutral_ground}), the feasible space would be limited to the power flow solutions. Since power flow is a feasibility problem, the objective function $f(\mathbf{x})$ can be a zero function or squared sum of all generator active powers. Authors in \cite{8442948} apply this approach to solve an AC power flow problem for purely AC systems by adding the constraints associated with generator powers and voltage control buses. The optimization-based approach to solving power flow (PF) is more flexible in handling various system constraints and is generally more robust against system limitations. It is relatively straightforward to incorporate any system limitations (e.g., voltage limits, generator reactive power limits) as constraints in the optimization model, compared to traditional Newton-Raphson methods \cite{852122}. Additionally, the optimization-based approach provides better assurance of solution feasibility, as the feasibility space can be controlled and confined using different constraints. However, the advantages of the optimization-based approach come at the cost of increased problem formulation complexity and higher computational costs compared to traditional power flow solution methods. 

\section{Test Case and Numerical Results}
 \subsection{System description}\label{Test_case}

    \begin{figure}[!tbp]
        \centering
        \includegraphics[width=\columnwidth]{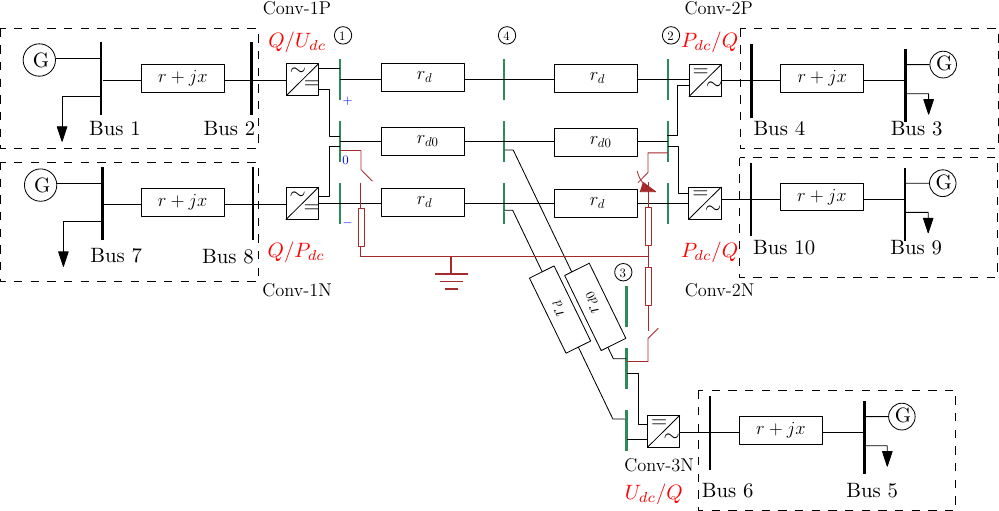}
        \caption{Test system with control modes representations for the Case-1}
        \label{fig: test_case_toycase_mcdc_PF_case_1}
    \end{figure}
    \begin{table}[!tbp]
    \scriptsize
\centering
\caption{Test case and control mode combinations}
\label{tab:cases}
\begin{tabular}{@{}cl|ccccc@{}}
\hline
 \multicolumn{2}{c}{Conv-} & 1P    & 2P    & 3N     & 1N    & 2N    \\ \hline
\multirow{2}{*}{Case-1} &$d$-axis              & $U^{dc}$   &$P^{ac}$   & $U^{dc}$   &$P^{ac}$   &$P^{ac}$   \\ 
                        &$q$-axis              & $Q$   & $Q$   & $Q$   & $Q$   & $Q$   \\ \hdashline 
                        & $d$-axis & $U^{dc}$ & $P^{ac}$                           & $U^{dc}$ & $P^{ac}$ & $P^{ac}$           \\
\multirow{-2}{*}{Case-2} & $q$-axis & $Q$ & \cellcolor[HTML]{D1F1DA}$U^{ac}$   & $Q$ & $Q$ & $Q$                           \\ \hdashline
                         & $d$-axis & $U^{dc}$ & \cellcolor[HTML]{D1F1DA}DC-droop & $U^{dc}$ & $P^{ac}$ & $P^{ac}$                           \\
\multirow{-2}{*}{Case-3} & $q$-axis & $Q$ & $Q$                           & $Q$ & $Q$ & $Q$                           \\ \hdashline
                         & $d$-axis & $U^{dc}$ & $P^{ac}$                           & $U^{dc}$ & $P^{ac}$ & $P^{ac}$                           \\
\multirow{-2}{*}{Case-4} & $q$-axis & $Q$ & \cellcolor[HTML]{D1F1DA}AC-droop & $Q$ & $Q$ & $Q$                           \\ \hdashline
                         & $d$-axis & $U^{dc}$ & $P^{ac}$                           & $U^{dc}$ & $P^{ac}$ & \cellcolor[HTML]{D1F1DA}DC-droop \\
\multirow{-2}{*}{Case-5} & $q$-axis & $Q$ & $Q$                           & $Q$ & $Q$ & $Q$   \\                        
                        \hline
\end{tabular}
\end{table}

 \begin{table}[!tbp]
    \scriptsize
    \centering
    \caption{Converter setpoints (in $pu$) for the five cases}
    \label{tab: converter_setpoints}
    % \resizebox{\columnwidth}{!}{
    % \begin{tabular}{@{}ccccccc@{}}
    \begin{tabular}{@{}lllllll@{}}
    \hline 
    \multicolumn{1}{l}{}    & \multicolumn{1}{l}{}    & 1P                          & 2P                                & 3N                            & 1N                              & 2N                                \\ \hline
    \multirow{2}{*}{Case-1} & $d$-axis                  & $U^{dc}$ = 1.0                   & $P^{ac}$ = - 0.76070                  & $U^{dc}$ = -1.001                  & $P^{ac}$ = 0.87193                  & $P^{ac}$ = - 0.42641                  \\
                            & $q$-axis                  & $Q$ = -0.2                  & $Q$ = 0.1                         & $Q$ = - 0.15                  & $Q$ = - 0.3                     & $Q$ = - 0.05                      \\ \hline
    \multirow{2}{*}{Case-2} & $d$-axis                  & $U^{dc}$ = 1.0                   & $P^{ac}$ = - 0.76070                  & $U^{dc}$ = -1.001                  & $P^{ac}$ = 0.87193                  & $P^{ac}$ = - 0.42641                  \\
                            & $q$-axis                  & $Q$ = -0.2                  & $U^{ac}$ = 1.05                        & $Q$ = - 0.15                  & $Q$ = - 0.3                     & $Q$ = - 0.05                      \\ \hline
    \multirow{4}{*}{Case-3} & \multirow{3}{*}{$d$-axis} & \multirow{3}{*}{$U_{dc }$= 1.0}  & $U^{dc}$ = 1.0                         & \multirow{3}{*}{$U_{dc }$= -1.001} & \multirow{3}{*}{$P^{ac}$ = 0.87193} & \multirow{3}{*}{$P^{ac}$ = - 0.42641} \\
                            &                         &                             & $P^{ac}$ = - 0.76070                  &                               &                                 &                                   \\
                            &                         &                             & $K_{droop}^{dc}$= 0.1                          &                               &                                 &                                   \\  \cdashline{3-4}
                            & $q$-axis                  & $Q$ = -0.2                  & $Q$ = 0.1                         & $Q$ = - 0.15                  & $Q$ = - 0.3                     & $Q$ = - 0.05                      \\ \hline
    \multirow{4}{*}{Case-4} & $d$-axis                  & $U^{dc}$ = 1.0                   & $P^{ac}$ = - 0.76070                  & $U^{dc}$ = -1.001                  & $P^{ac}$ = 0.87193                  & $P^{ac}$ = - 0.42641                  \\ \cdashline{3-4}
                            & \multirow{3}{*}{$q$-axis} & \multirow{3}{*}{$Q$ = -0.2} & $U^{ac}$ = 1.05                        & \multirow{3}{*}{$Q$ = - 0.15} & \multirow{3}{*}{$Q$ = - 0.3}    & \multirow{3}{*}{$Q$ = - 0.05}     \\
                            &                         &                             & $Q$ = 0.1                         &                               &                                 &                                   \\
                            &                         &                             & $K_{droop}^{ac}$= 0.05                         &                               &                                 &                                   \\ \hline
    \multirow{4}{*}{Case-5} & \multirow{3}{*}{$d$-axis} & \multirow{3}{*}{$U^{dc}$= 1.0}  & \multirow{3}{*}{$P^{ac}$ = - 0.76070} & \multirow{3}{*}{$U_{dc }$= -1.001} & \multirow{3}{*}{$P^{ac}$ = 0.87193} & $U^{dc}$ = -1.0                         \\
                            &                         &                             &                                   &                               &                                 & $P^{ac}$= 0.42641                     \\
                            &                         &                             &                                   &                               &                                 & $K_{droop}^{dc}$ = 0.1                         \\ \cdashline{6-7}
                            & $q$-axis                  & $Q$ = -0.2                  & $Q$ = 0.1                         & $Q$ = - 0.15                  & $Q$ = - 0.3                     & $Q$ = - 0.05   \\ \hline                 
    \end{tabular}
    % }
    \end{table}

    The test system used to demonstrate the functionalities of the model is presented in \cref{fig: test_case_toycase_mcdc_PF_case_1}. This system consists of a multi-terminal bipolar HVDC network connecting 5 asynchronous AC zones. Here, the DC network consists of two bipolar and one monopolar converter station, as well as DC branches in a four-bus (DC) system. This system can be seen as an asymmetric monopolar tapping on a bipolar link or it can also be seen as a bipolar system operating in unbalanced mode when one of the converter poles (connected to DC bus \circled{3} positive terminal, not displayed in the figure) is out of service.
    
    As explained in \cref{sec: AC-droop}, the power flow formulation would require voltage references for positive, negative, and neutral voltage levels. The neutral terminal reference is provided by grounding the neutral terminal of DC bus \circled{2}. Whereas the positive and negative voltage references are provided by the converter in DC voltage control mode ($U^{dc}$ control or DC-droop). Since there can be various combinations of the control choices, five cases, as defined in \cref{tab:cases}, are chosen for the demonstration.
    % Case-1 with conv-1N out and

    It can be observed that the $d$-axis of at least one converter per pole is either in $U^{dc}$ or $droop$ control mode to ensure control over the DC voltage. In this system (\cref{fig: test_case_toycase_mcdc_PF_case_1}) two converters are connected to the positive pole and three to the negative pole. Case-1 considers Conv-1P and 3N in DC voltage control mode ($U^{dc}$), while the remaining converters are operated in active power control mode ($P^{ac}$). For the $q$-axis, all of the converters are in reactive power control mode ($Q$).
    
     Cases- 2 to 5 have a change of one converter control mode w.r.t. Case-1, as highlighted in \cref{tab:cases} with a green color. In Case-2, $q$-axis control of Conv-2P is changed to $U^{ac}$ control mode, thus, the converter should adjust its reactive power to maintain the desired voltage magnitude at AC bus 4. In Case-3, everything is the same except the $d$-axis control mode of Conv-2P, which is changed to DC-droop i.e., a droop relation between the DC bus voltage (positive terminal of DC bus 2) and the converter active power. Similarly, the control mode changes in Case-4 and Case-5 each can be observed in \cref{tab:cases}. 

     For all five cases, the AC grid controls, i.e., generator controls are kept the same. Since all five generators are in five different synchronous areas, each of them acts as a reference bus in their respective system. Thus, for every generator bus, the voltage magnitude is set at 1.0\, $pu$, and the voltage angle at 0 $rad$. Whereas for the DC system, control setpoints are set as reported in \cref{tab: converter_setpoints}.

     \subsection{Validation of the model}
    Since there are no pre-existing benchmark cases for the unbalanced HVDC systems, a PSCAD model is developed for the test case presented in \cref{fig: test_case_toycase_mcdc_PF_case_1}. For this PSCAD model all control modes corresponding to all five cases in \cref{tab: converter_setpoints} are implemented. The results of the proposed power flow model presented in this paper, are validated by comparing its results with the steady-state values obtained from the PSCAD simulations. Table \ref{tab: dcvoltages_case_1} compares DC bus terminal voltages. Here one control combination (Case-3) from \cref{tab: converter_setpoints} is presented for comparison.
    It is observed that DC bus voltages obtained from the presented power flow models match with those from the PSCAD simulation up to the 4$^{th}$ decimal digit. The maximum difference observed in the DC terminal voltages is 0.000014\,$pu$. Similarly, \cref{tab: conv_powers_case_1} compares converter active and reactive power flows, which matches up to the 3$^{rd}$ decimal digit and the maximum difference observed is 0.00057\,$pu$. Finally, the AC bus voltage magnitudes and angles (in $rad$) are compared. The maximum difference observed for the AC bus voltage magnitude is 0.0012\, $pu$, which is acceptable in literature for a PSCAD-based validation, particularly for AC network quantities \cite{kim2024solving}. It is important to mention that PSCAD simulations do not provide a constant value at steady-state but a waveform with some fluctuations, therefore, average values are chosen for the comparison. For a more precise assessment of the AC side voltages, all the two bus AC systems are solved manually for the active and reactive power values in \cref{tab: conv_powers_case_1}, and the results are found to match with the proposed model up to 5$^{th}$ decimal digit for AC bus voltage magnitude and angle. Similarly, results for Case-2 to Case-5 are also validated against the PSCAD simulations and found to have a similar accuracy. Thus, the proposed model is validated with a PSCAD simulation.
    \begin{table*}[!tbp]
    \scriptsize
    \centering
    \caption{DC bus voltage magnitudes for Case-3}
    \label{tab: dcvoltages_case_1}
    \resizebox{\columnwidth}{!}{
    \begin{tabular}{@{}ccrrrrrrrrr@{}}
    \hline
    \multicolumn{1}{l}{}    & \multicolumn{1}{l}{}      & \multicolumn{3}{c}{Propsed model}                                                         & \multicolumn{3}{c}{PSCAD simulation}                                                      & \multicolumn{3}{c}{Difference}                                                                 \\ \hline
    \multicolumn{1}{l}{}    & \multicolumn{1}{l}{busdc} & \multicolumn{1}{l}{Positive} & \multicolumn{1}{l}{Negative} & \multicolumn{1}{l}{Neutral} & \multicolumn{1}{l}{Positive} & \multicolumn{1}{l}{Negative} & \multicolumn{1}{l}{Neutral} & \multicolumn{1}{l}{Positive} & \multicolumn{1}{l}{Negative} & \multicolumn{1}{l}{Neutral} \\ \hline
                            & 1                    & 1.000188             & -1.022990 & 0.000188  & 1.000190             & -1.022980 & 0.000183   & 0.000002  & 0.000010  & -0.000006 \\
    \multicolumn{1}{c}{Case-3}      & 2                    & 0.985679             & -1.008292 & 0.000000  & 0.985680             & -1.008280 & 0.000000   & 0.000001  & 0.000012  & 0.000000  \\
    \multicolumn{1}{c}{(DC-droop)}  & 3                    & \multicolumn{1}{l}{} & -1.008273 & -0.007273 & \multicolumn{1}{l}{} & -1.008270 & -0.007270  & 0.000000  & 0.000003  & 0.000003  \\
                            & 4                    & 0.992934             & -1.013185 & -0.002362 & 0.992920             & -1.013180 & -0.002362  & -0.000014 & 0.000005  & 0.000000  \\ 
    \hline
    \end{tabular}
    }
    \end{table*}
   
    \begin{table}[!tbp]
    \scriptsize
    \centering
    \caption{Active and reactive power injections at the PCC for Case-3}
    \label{tab: conv_powers_case_1}
    \begin{tabular}{@{}c|rr|rr|rr@{}}
    \hline
    \multicolumn{1}{c|}{} & \multicolumn{2}{c}{Proposed model}               & \multicolumn{2}{c}{PSCAD simulation}            & \multicolumn{2}{c}{Difference}                                                               \\
    \multicolumn{1}{c|}{Conv}                 & \multicolumn{1}{c}{$P^{ac}$} & \multicolumn{1}{c}{$Q^{ac}$} & \multicolumn{1}{c}{$P^{ac}$} & \multicolumn{1}{c}{$Q^{ac}$} & \multicolumn{1}{c}{$P^{ac}$} & \multicolumn{1}{c}{$Q^{ac}$} \\
    \hline
    1P                   & 0.62986                & -0.20000               & 0.62970                & -0.20000               & -0.00016               & 0.00000                \\
    2P                   & -0.61749               & 0.10000                & -0.61748               & 0.10000                & 0.00001                & 0.00000                \\
    3N                   & -0.42488               & -0.15000               & -0.42545               & -0.15000               & -0.00057               & 0.00000                \\
    1N                   & 0.87193                & -0.30000               & 0.87195                & -0.30000               & 0.00002                & 0.00000                \\
    2N                   & -0.42641               & -0.05000               & -0.42644               & -0.05000               & -0.00003               & 0.00000                \\
    
    \hline
    \end{tabular}
    \end{table}

\subsection{Numerical results for different control modes}

\begin{figure}[!htb]
    \centering
    \includegraphics[width= 0.8\columnwidth]{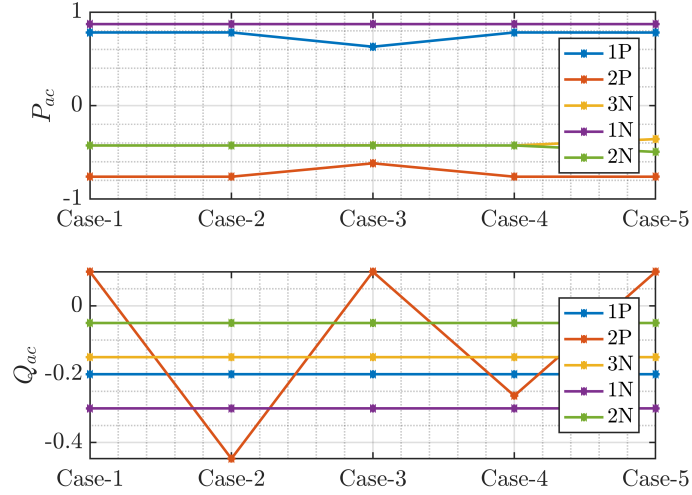}
    \caption{Converter power flows for the different control modes}
    \label{fig: conv_power_base}
\end{figure}

The power flow problem is solved for all five control modes defined in \cref{tab:cases} with the setpoints in \cref{tab: converter_setpoints}. Active and reactive powers of converters are plotted in \cref{fig: conv_power_base}. It can be observed that the solutions obtained from the proposed power flow model comply with predefined setpoints. 

The $q$-axis control for all converters, except Conv-2P, is $Q$ control, therefore, their reactive power flows remain the same in all five cases, as seen in the lower plot of \cref{fig: conv_power_base}. For Conv-2P, the reactive power can be seen changing according to its control mode. In Case-2, Conv-2P operates in $U^{ac}$ control mode to maintain the magnitude of AC bus 4 at 1.05 $pu$, which was around 0.99 $pu$ in Case-1. Therefore, the reactive power ($Q$) of Conv-2P drops from 0.1 $pu$ to -0.44 $pu$, i.e., now the converter injects reactive power into the system. In Case-4, Conv-2P operates in $U^{ac}$-$Q$ droop, i.e., AC-droop control mode; therefore, the drop in reactive power is less (-0.26 $pu$).

\begin{figure}[!htb]
    \centering
    \includegraphics[width= 0.8\columnwidth]{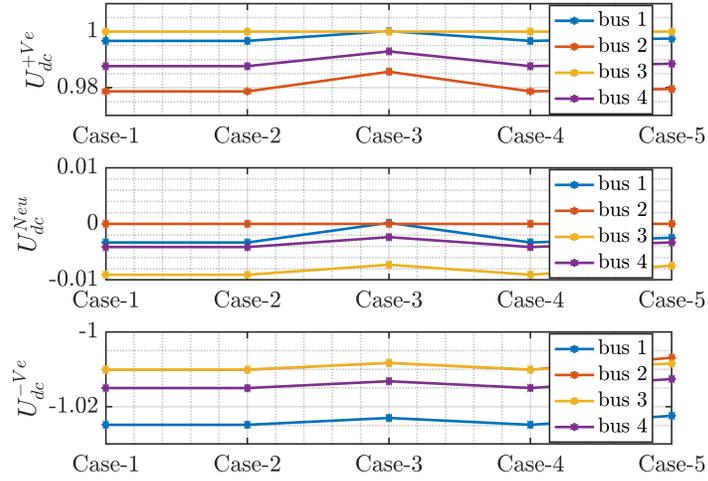}
    \caption{DC bus voltages for the different control modes}
    \label{fig: PM_dcbus_voltage_base}
\end{figure}
On the other hand, the active power flows of the converter remain (largely) unchanged in Case-2 and Case-4. These flows change for Case-3 and Case-5, i.e., DC-droop control modes. In Case-3, the DC-droop control is applied to Conv-2P, and therefore, its active power flow is changed. As a result, the active power flow of the slack converter of the positive layer, i.e., Conv-1P, also gets adjusted, as seen in \cref{tab: conv_power_outage}. Similarly, in Case-5, the DC-droop control is applied to Conv-2N, therefore, active power flows of Conv-2N and Conv-3N are changed. This change in active power flow also affects the DC bus voltages, as seen in \cref{fig: PM_dcbus_voltage_base}. It is important to note that the magnitude of all three voltage polarities, i.e., positive, negative, and neutral, is changed for both Case-3 and Case-5. Thus, a change in the active power flow of the positive layer may have little impact on the active power flows of the negative layer converters, but the voltage profiles of all three polarities are affected to the same degree (\cref{fig: PM_dcbus_voltage_base}). Further, it can also be seen that the DC grid voltages do not change, as expected, for change in the $q$-axis control mode, i.e., Case-1, Case-2, and Case-4. 

\subsection{System states following an outage of Conv-1N}

This section presents the results of power flow after the outage of Conv-1N. Since control modes of the converter in the system would govern the states of the system following the outage, the impact of the outage is studied for all five cases at their respective setpoints as mentioned in \cref{tab: converter_setpoints}. System states before and after the outage are compared.

\begin{figure}[tbh!]
    \centering
    \includegraphics[width= 0.8\columnwidth]{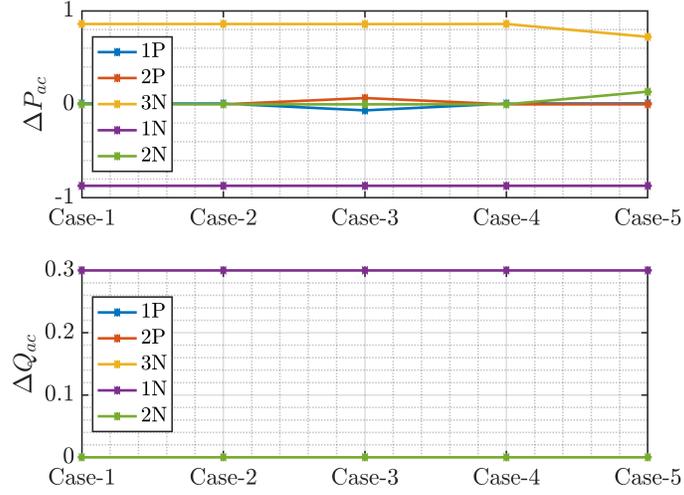}
    \caption{Deviation in converter power flows after outage of Conv-1N}
    \label{fig: conv_power_deviations_outage}
\end{figure}

\cref{tab: conv_power_outage} presents active and reactive power flows for each converter before and after the outage of Conv-1N. The change in quantities following an outage is represented as a deviation, which is defined as the value after the outage minus the value before the outage. For example, the deviation in active power at converter Conv-2P is calculated as \( \Delta P_{ac, 2P} = P_{ac, 2P}^{\text{after outage}} - P_{ac, 2P}^{\text{before outage}} \). \cref{fig: conv_power_deviations_outage} presents the deviations for active and reactive power flows. Since Conv-1N operates in $P^{ac}$ and $Q$ control modes for all cases before its outage, its power deviation is the same in all the cases, i.e., equal to before the outage operating values. Given that the converter loss occurs at the negative pole, the resulting active power imbalance must be compensated by other converters on this pole. For Case-1 to Case-4, Conv-2N operates in $P^{ac}$ control mode, i.e., at constant active power. Therefore, the entire loss of active power is compensated by Conv-3N, which operates as a slack converter. It can be observed from \cref{tab: conv_power_outage} and \cref{fig: conv_power_deviations_outage}, that the power deviation of Conv-3N is nearly equal in magnitude (with slight differences due to variations in losses) but opposite in sign to the deviation of the outgoing Conv-1N, for Case-1 to Case-4. Whereas for Case-5, Conv-2N operates in DC-droop control mode, thus adjusting the active power output according to the DC voltage deviations. Therefore, the loss of power balance in the negative layer caused by the outage is shared by both Conv-2N and Conv-3N, thus reducing the burden on just the slack converter, as seen in \cref{fig: conv_power_deviations_outage}. The power-sharing between converters and, thus, limiting the DC voltage deviation can be optimized according to system operation requirements, which is out of the scope of this paper. However, it is important to note that the active power flows of the converter also change, although the degree of deviation is less. This slight deviation can be significant under operations near system boundaries, implying that the outage of a converter in one layer (or pole) also affects the flows in the other layer (or pole), and the deviation depends on the control modes of the converters. It is further observed in Case-3 where Conv-2P of the positive layer operates in DC-droop control mode, the active power flows of both Conv-1P and Conv-2P face deviations. These cross-layer deviations in power flows occur due to the voltage deviations caused by the common neutral.

\begin{figure}[tbh!]
    \centering
    \includegraphics[width= 0.8\columnwidth]{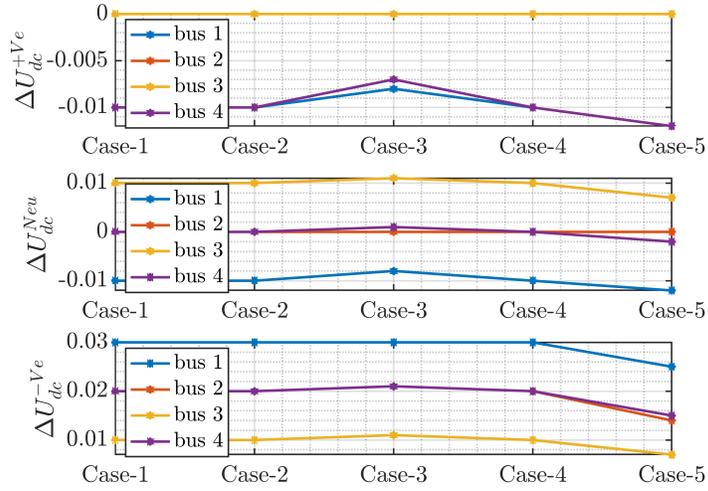}
    \caption{Deviation in DC bus voltages after outage of Conv-1N}
    \label{fig: dcvoltage_deviations_outage}
\end{figure}
\vspace{0.05cm}
\begin{table}[htp]
\centering
\scriptsize
\caption{Converter power flows before and after outage of Conv-1N}
\label{tab: conv_power_outage}
\begin{tabular}{@{}ccrrrr@{}}
\hline 
\multicolumn{1}{l}{}    & \multicolumn{1}{l}{} & \multicolumn{2}{c}{Before outage}                   & \multicolumn{2}{c}{After outage}                \\ \hline
\multicolumn{1}{l}{}    & Conv                 & \multicolumn{1}{c}{$P^{ac}$} & \multicolumn{1}{c}{$Q$} & \multicolumn{1}{c}{$P^{ac}$} & \multicolumn{1}{c}{$Q$} \\ \hline
\multirow{5}{*}{Case-1} & 1P                   & 0.7824                 & -0.2000                & 0.7906                 & -0.2000                \\
                        & 2P                   & -0.7607                & 0.1000                 & -0.7607                & 0.1000                 \\
                        & 3N                   & -0.4270                & -0.1500                & 0.4327                 & -0.1500                \\
                        & 1N                   & 0.8719                 & -0.3000                & 0.0000                 & 0.0000                 \\
                        & 2N                   & -0.4264                & -0.0500                & -0.4264                & -0.0500                \\ \cdashline{2-6}
\multirow{5}{*}{Case-2} & 1P                   & 0.7829                 & -0.2000                & 0.7911                 & -0.2000                \\
                        & 2P                   & -0.7607                & -0.4473                & -0.7607                & -0.4473                \\
                        & 3N                   & -0.4270                & -0.1500                & 0.4327                 & -0.1500                \\
                        & 1N                   & 0.8719                 & -0.3000                & 0.0000                 & 0.0000                 \\
                        & 2N                   & -0.4264                & -0.0500                & -0.4264                & -0.0500                \\ \cdashline{2-6}
\multirow{5}{*}{Case-3} & 1P                   & 0.6299                 & -0.2000                & 0.5649                 & -0.2000                \\
                        & 2P                   & -0.6175                & 0.1000                 & -0.5503                & 0.1000                 \\
                        & 3N                   & -0.4249                & -0.1500                & 0.4338                 & -0.1500                \\
                        & 1N                   & 0.8719                 & -0.3000                & 0.0000                 & 0.0000                 \\
                        & 2N                   & -0.4264                & -0.0500                & -0.4264                & -0.0500                \\ \cdashline{2-6}
\multirow{5}{*}{Case-4} & 1P                   & 0.7825                 & -0.2000                & 0.7906                 & -0.2000                \\
                        & 2P                   & -0.7607                & -0.2630                & -0.7607                & -0.2630                \\
                        & 3N                   & -0.4270                & -0.1500                & 0.4327                 & -0.1500                \\
                        & 1N                   & 0.8719                 & -0.3000                & 0.0000                 & 0.0000                 \\
                        & 2N                   & -0.4264                & -0.0500                & -0.4264                & -0.0500                \\ \cdashline{2-6}
\multirow{5}{*}{Case-5} & 1P                   & 0.7817                 & -0.2000                & 0.7913                 & -0.2000                \\
                        & 2P                   & -0.7607                & 0.1000                 & -0.7607                & 0.1000                 \\
                        & 3N                   & -0.3584                & -0.1500                & 0.3638                 & -0.1500                \\
                        & 1N                   & 0.8719                 & -0.3000                & 0.0000                 & 0.0000                 \\
                        & 2N                   & -0.4954                & -0.0500                & -0.3599                & -0.0500   \\            \hline 
\end{tabular}
\end{table}

% \clearpage
\begin{table}[htp]
\centering
\scriptsize
\caption{DC bus voltages before and after outage of Conv-1N}
\label{tab: dcbus_volatges_outage}
 % \resizebox{\columnwidth}{!}{
\begin{tabular}{@{}cccccccc@{}}
\hline 
\multicolumn{1}{l}{}    & \multicolumn{1}{l}{} & \multicolumn{3}{c}{Before outage}              & \multicolumn{3}{c}{After outage}          \\ 
\hline
\multicolumn{1}{l}{Case-}    & DCbus                & Positive             & Negative & Neutral  & Positive             & Negative & Neutral  \\
\hline
\multirow{4}{*}{1} & 1                    & 0.99669              & -1.02482 & -0.00331 & 0.98678              & -0.99515 & -0.01322 \\
                        & 2                    & 0.97868              & -1.01011 & 0.00000  & 0.96858              & -0.99017 & 0.00000  \\
                        & 3                    & \multicolumn{1}{l}{} & -1.01006 & -0.00906 & \multicolumn{1}{l}{} & -1.00014 & 0.00086  \\
                        & 4                    & 0.98769              & -1.01500 & -0.00412 & 0.97768              & -0.99515 & -0.00412 \\
                        \cdashline{2-8}
\multirow{4}{*}{2} & 1                    & 0.99668              & -1.02482 & -0.00332 & 0.98677              & -0.99516 & -0.01323 \\
                        & 2                    & 0.97866              & -1.01012 & 0.00000  & 0.96856              & -0.99018 & 0.00000  \\
                        & 3                    & \multicolumn{1}{l}{} & -1.01006 & -0.00906 & \multicolumn{1}{l}{} & -1.00014 & 0.00086  \\
                        & 4                    & 0.98767              & -1.01500 & -0.00413 & 0.97766              & -0.99516 & -0.00412 \\
                        \cdashline{2-8}
\multirow{4}{*}{3} & 1                    & 1.00019              & -1.02299 & 0.00019  & 0.99198              & -0.99252 & -0.00802 \\
                        & 2                    & 0.98568              & -1.00829 & 0.00000  & 0.97896              & -0.98752 & 0.00000  \\
                        & 3                    & \multicolumn{1}{l}{} & -1.00827 & -0.00727 & \multicolumn{1}{l}{} & -0.99752 & 0.00348  \\
                        & 4                    & 0.99293              & -1.01319 & -0.00236 & 0.98547              & -0.99252 & -0.00151 \\
                        \cdashline{2-8}
\multirow{4}{*}{4} & 1                    & 0.99669              & -1.02482 & -0.00331 & 0.98678              & -0.99515 & -0.01322 \\
                        & 2                    & 0.97868              & -1.01011 & 0.00000  & 0.96858              & -0.99017 & 0.00000  \\
                        & 3                    & \multicolumn{1}{l}{} & -1.01006 & -0.00906 & \multicolumn{1}{l}{} & -1.00014 & 0.00086  \\
                        & 4                    & 0.98768              & -1.01500 & -0.00412 & 0.97768              & -0.99515 & -0.00412 \\
                        \cdashline{2-8}
\multirow{4}{*}{5} & 1                    & 0.99753              & -1.02242 & -0.00247 & 0.98597              & -0.99754 & -0.01403 \\
                        & 2                    & 0.97954              & -1.00690 & 0.00000  & 0.96776              & -0.99335 & 0.00000  \\
                        & 3                    & \multicolumn{1}{l}{} & -1.00845 & -0.00745 & \multicolumn{1}{l}{} & -1.00173 & -0.00073 \\
                        & 4                    & 0.98853              & -1.01259 & -0.00330 & 0.97687              & -0.99754 & -0.00492 \\
                        \hline 
\end{tabular}
% }
\end{table}

\cref{tab: dcbus_volatges_outage} presents DC bus voltages at each terminal before and after the converter outage, and the DC voltage deviation is plotted in \cref{fig: dcvoltage_deviations_outage}. Following the outage in the negative pole, the power flow redistribution happens, and thus the terminal voltage changes, according to the control mode of the converters. As seen in \cref{fig: dcvoltage_deviations_outage}, voltages of positive terminals of all DC buses change for all the cases (DC bus 3 positive terminal is isolated and therefore shown as zero deviation). For this test case and operating modes, the deviation for the positive pole is negative, i.e., the overall voltage level of the positive pole drops slightly. Whereas the deviation for the negative pole is positive, i.e., the voltage level of negative terminals of all the DC buses increases. An increase in negative voltage is a decrease in its absolute value, which is an expected system behavior when a converter feeding power into the DC system goes out. The exact degree of deviation depends on the control modes. As in Case-5, Conv-2N operates in DC-droop control mode and shares the power imbalance, it also reduces the maximum voltage deviation from 0.03 $pu$ to 0.025 $pu$, as seen in \cref{fig: dcvoltage_deviations_outage}. For the neutral terminals, both positive and negative deviations occur, i.e., both voltage rise and fall occur at different buses. 

Thus, an outage of a converter in one pole of a bipolar converter station affects the power flows and voltage levels in the entire system. Furthermore, the impact of a contingency on system states depends not only on the control modes of the converters in the pole of contingency but also on the control modes of the converters in the other pole. These results have also been validated with PSCAD simulations.

\subsection{AC-droop control}

\begin{figure}[bh]
    \centering
    \includegraphics[width= \columnwidth]{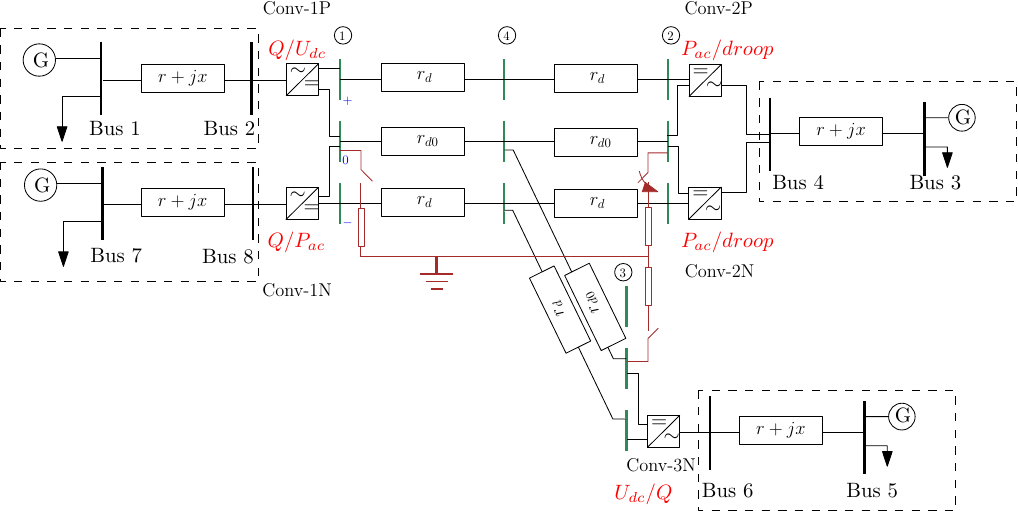}
    \caption{Test systems with both Conv-2P and Conv-2N connected to the same AC bus and operating in AC-droop control mode}
    \label{fig: test_case_toycase_mcdc_PF_AC-droop}
\end{figure}

To resemble a bipolar converter station with both converter poles connected to the same AC bus, the test case in \cref{fig: test_case_toycase_mcdc_PF_case_1}, is modified by connecting both Conv-2P and Conv-2N to the same AC bus, as in \cref{fig: test_case_toycase_mcdc_PF_AC-droop}. Both the converters at AC bus 4 can be controlled independently for both $d$-axis and $q$-axis control. To analyze the impact of $q$-axis control modes on reactive power from Conv-2P and Conv-2N, three different control cases are considered as seen in \cref{tab: AC-droop-cases}. All other converter control modes and their setpoints are the same as Case-1 in \cref{tab: converter_setpoints}. 

\begin{table}[thb]
\centering
\scriptsize
\caption{Test cases for AC-droop control mode}
\label{tab: AC-droop-cases}
\begin{tabular}{@{}cllll@{}} 
\hline
\multicolumn{1}{l}{}    &         & Case-A                      & Case-B                      & Case-C                      \\  \hline
\multirow{2}{*}{$d$-axis} & Conv-2P & $P^{ac}$                         & $P^{ac}$                         & $P^{ac}$                         \\
                        & Conv-2N & $P^{ac}$                         & $P^{ac}$                         & $P^{ac}$                         \\
\multirow{2}{*}{$q$-axis} & Conv-2P & $Q$                         & $U^{ac}$                         & AC-droop                    \\
                        & Conv-2N & $Q$                         & $Q$                         & AC-droop                    \\ \hline
$U^{ac}$ ($pu$)                    & AC bus-4 & \multicolumn{1}{c}{0.99930} & \multicolumn{1}{c}{1.05000} & \multicolumn{1}{c}{1.04014} \\ \cdashline{2-5}
\multirow{2}{*}{$Q$ ($pu$)}    & Conv-2P & \multicolumn{1}{c}{0.1000}  & \multicolumn{1}{c}{-0.3947} & \multicolumn{1}{c}{-0.0973} \\
                        & Conv-2N & \multicolumn{1}{c}{-0.0500} & \multicolumn{1}{c}{-0.0500} & \multicolumn{1}{c}{-0.2473} \\ \hline
\end{tabular}
\end{table}

\begin{figure}[!htb]
    \centering
    \includegraphics[width= 0.8\columnwidth]{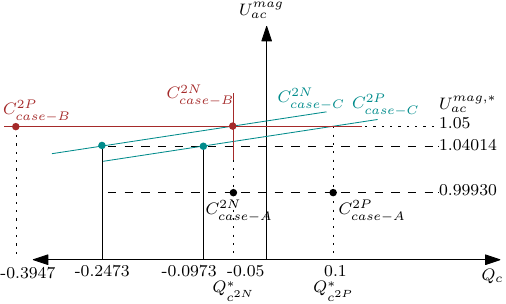}
    \caption{Reactive power sharing between Conv-2P and Conv-2N with AC-droop ($U^{ac}$ - $Q_{c}$) control mode}
    \label{fig: num_results_ac_droop-droop}
\end{figure}

The power flow problem is solved for Case-A, Case-B, and Case-C in \cref{tab: AC-droop-cases}. Numerical results for the voltage magnitude at AC bus 4, as well as the reactive powers of Conv-2P and Conv-2N, are presented in \cref{tab: AC-droop-cases}. Since the DC grid and other AC systems are largely unaffected by these changes, their values are not reported here, as they are equal to the values in Case-1.

\cref{fig: num_results_ac_droop-droop} shows a graphical representation of the  $q$-axis control modes and operating points (marked dots) in each of the three cases. In Case-A, when both  Conv-2P and Conv-2N  operate in $Q$ control mode, their reactive powers are fixed at the respective reference values and the resultant magnitude of the bus 4 voltage is 0.99930 $pu$. In case B, Conv-2P is changed to $U^{ac}$ control mode with reference equal to 1.05\,$pu$, while Conv-2N remains in $Q$ control. The complete burden of raising the voltage falls on Conv-2P and it supplies the entire additional need of reactive power. Its reactive power withdrawal at the PCC changes from 0.1 to -0.3947 $pu$ i.e., instead of absorbing 0.1 $pu$ now it supplies -0.3947 $pu$, while Conv-2N continues to withdraw 0.05 $pu$. In Case-C, the $q$-axis control mode of both Conv-2P and Conv-2N is changed to operate in $U^{ac}$- $Q$ droop, i.e., AC-droop control mode. In this case, both converters share the reactive power needed to raise the AC grid voltage (at AC bus 4). As a result, the reactive power of Conv-2P changes from 0.1 to -0.0973, i.e., a drop of 0.1973, whereas Conv-2N also shares the same amount, i.e., a drop of 0.1973 $pu$, changing from -0.05 to -0.2473. Since both the converters have the same droop coefficients ($K_{droop}^{ac}$=0.05), they share the change in reactive power equally. However, these droop coefficients can be adjusted as per the system requirements. Such a droop control mode can be advantageous in an unbalanced bipolar system. In this scenario, a converter operating with lower active power, thus having a higher margin for reactive power can be assigned to have a lower droop coefficient resulting in a higher share of reactive power.

\subsection{Large test cases}
The proposed model is also successfully tested for larger test cases, up to 3120 bus systems, that can be validated at \cite{PMMCDC.jl}, the open-source tool in the Julia/JuMP framework.

\section{Conclusion}
This paper presents a unified power flow model for the unbalanced operation of bipolar HVDC grids in hybrid AC/DC systems. The proposed power flow model represents the DC grid with a multiconductor model with an explicit representation of each terminal at the DC bus and each conductor of a DC branch. Further, with an explicit representation of both converters at a bipolar converter station, this model incorporates all basic control modes of VSC HVDC systems in balanced HVDC grids and extends them to bipolar HVDC grids with unbalanced operation. Additionally, an AC-droop control mode is modeled. The AC-droop control mode allows reactive power sharing among multiple HVDC converters connected to the same AC bus. Thus, this control mode can be highly useful in unbalanced operation where a converter operating at lower active power can be assigned to provide higher reactive power support by adjusting its droop coefficient accordingly. 
The proposed model is demonstrated on a five-terminal MTDC system with an unbalanced DC grid. Analysis of different control mode combinations shows that any change of active power in one pole (either positive or negative) may have little impact on the power flows of the other pole (or layer) but it would affect the voltages of all three voltage polarities to a similar degree. However, if this change of voltage further influences the power output of a converter in DC-droop control, it might bring bigger changes in the power flow, depending upon the control setting. Therefore, a change in the power flow (or outage) in one pole can potentially cause operating and equipment limit violations in the other pole. The power flow model presented in this paper accurately captures such details for bipolar HVDC grids under balanced and unbalanced operating conditions. 

\section*{Acknowledgement}
This work was supported by the Belgian Energy Transition Fund, FOD Economy, project DIRECTIONS. Special thanks to Dr. Frederik Geth for his valuable input. 